\documentclass[aps,prl,twocolumn,groupedaddress]{revtex4}

\newcommand{\mub}{{$\mu_{B}$ }}
\newcommand{\mubp}{{$\mu_{B}$}}

\newcommand{\ab}{\textit{ab initio }}

\newcommand{\Ca}{CaMnO$_3$ }

\newcommand{\La}{LaMnO$_3$ }

\newcommand{\ABfive}{A$_{0.5}$B$_{0.5}$MnO$_3$ }

\newcommand{\LaCax}{La$_{1-x}$Ca$_x$MnO$_3$ }

\newcommand{\LaCa}{La$_{0.5}$Ca$_{0.5}$MnO$_3$ }

\newcommand{\PrCa}{Pr$_{0.5}$Ca$_{0.5}$MnO$_3$ }
\newcommand{\PrCasix}{Pr$_{0.60}$Ca$_{0.40}$MnO$_3$ }

\newcommand{\PrCasixp}{Pr$_{0.60}$Ca$_{0.40}$MnO$_3$}

\newcommand{\LaCap}{La$_{0.5}$Ca$_{0.5}$MnO$_3$}

\newcommand{\Mnoct}{MnO$_{6}$ }

\newcommand{\pbnm}{\textit{Pbnm }}

\newcommand{\pntom}{\textit{Pn}2$_{1}$\textit{m }}

\newcommand{\ptonm}{\textit{P}2$_{1}$\textit{nm }}
\newcommand{\ptonmp}{\textit{P}2$_{1}$\textit{nm}}
\newcommand{\ptonb}{\textit{P}2$_{1}$\textit{nb }}

\newcommand{\ptom}{\textit{P}112$_{1}$/\textit{m }}
\newcommand{\ptomp}{\textit{P}112$_{1}$/\textit{m}}

\newcommand{\ptobp}{\textit{P}112$_{1}$/\textit{b}}
\newcommand{\poto}{\textit{P}12$_{1}$1 }

\newcommand{\pem}{\textit{P}11\textit{m }}
\newcommand{\pemp}{\textit{P}11\textit{m}}

\newcommand{\pbp}{\textit{P}11\textit{b}}

\newcommand{\eg}{e$_g$ }

\newcommand{\dxxp}{d$_{x^{2}-y^{2}}$}
\newcommand{\dyy}{d$_{3y^{2}-r^{2}}$ }

\newcommand{\dx}{d$_{3x^{2}-r^{2}}$ }
\newcommand{\dy}{d$_{3y^{2}-r^{2}}$ }
\newcommand{\ooneion}{O$^{-}$ }
\newcommand{\ooneionp}{O$^{-}$}
\newcommand{\oion}{O$^{2-}$ }

\newcommand{\mmm}{Mn$^{3+}$ }

\newcommand{\mmmm}{Mn$^{4+}$ }

\newcommand{\df}{d$^{4}$ }
\newcommand{\dfp}{d$^{4}$}

\usepackage{graphicx}
\usepackage{latexsym}

\begin{document}

%\preprint{}

\title{Competing crystal structures in \LaCap: conventional charge order versus Zener polarons }



\author{C.H. Patterson}

\affiliation{ Department of Physics and Centre for Scientific Computation,\\
University of Dublin, Trinity College, Dublin 2, Ireland}

\date{\today}

\begin{abstract}
Equilibrium crystal structures for \LaCa have been calculated using hybrid exact exchange and density functional methods. Two distinct ground states with either conventional checkerboard charge ordering or Zener polaron formation are found depending on the proportion of exact exchange used. The checkerboard state has mixed \dxxp, \dx and \dyy orbital ordering and CE-type magnetic order while the Zener polaron state has \dx and \dy ordering and A-type magnetic order. \end{abstract}



\maketitle


\section{Introduction}

Low temperature charge ordering (CO) transitions in manganites such as \LaCax are generic for doping in the range 0.4 $<$ x $<$ 0.9 \cite{Chen96, Radaelli97, Radaelli99, Li01, Nagai02, Kajimoto02, Loudon02, Pissas02}. They are observed as commensurate or incommensurate changes in unit cell dimension parallel to the crystallographic $\bf{b}$ axis \cite{Loudon04a}. Commensurate structures are found when x is a rational fraction such as 1/2, 2/3 or 3/4. Half-doped manganites, \ABfive with A $=$ La, Nd, Pr and B $=$ Ca, Sr, have been widely studied as the CO phase is believed to consist of a checkerboard (CB) pattern of \mmm and \mmmm ions \cite{Wollan55, Goodenough55, Radaelli97} in which \mmm ions are Jahn-Teller (JT) distorted while \mmmm ions are not. However, there is evidence from Hartree-Fock calculations \cite{Zheng03,Ferrari03} on \LaCa for a Zener polaron (ZP) electronic structure in which $\textit{all}$ Mn ions have a valence of 3.5 and a recent single-crystal neutron scattering study found a ZP crystal structure for \PrCasix \cite{Daoud02} where all Mn ions have an intermediate JT distortion. There is conflicting experimental evidence from resonant XRD experiments \cite{Grenier04} for a CB CO pattern in \PrCasix and a recent refinement of the structure of \PrCa using high resolution x-ray and neutron powder data favored a CB structure \cite{Goff04}. Both CB and ZP phases have been proposed to exist in a phase diagram for half-doped manganites in which the tolerance ratio for the rare earth and alkaline earth ions is varied \cite{Rivadulla02}. Powder neutron and x-ray diffraction experiments \cite{Radaelli97} and x-ray absorption near edge structure (XANES) \cite{Subias97, Garcia01} and resonant x-ray diffraction (RXD) \cite{Garcia01a} support a CB CO picture for \LaCap.

\begin{figure}[ht!]
\includegraphics[width=8.00cm,height=8.72cm]{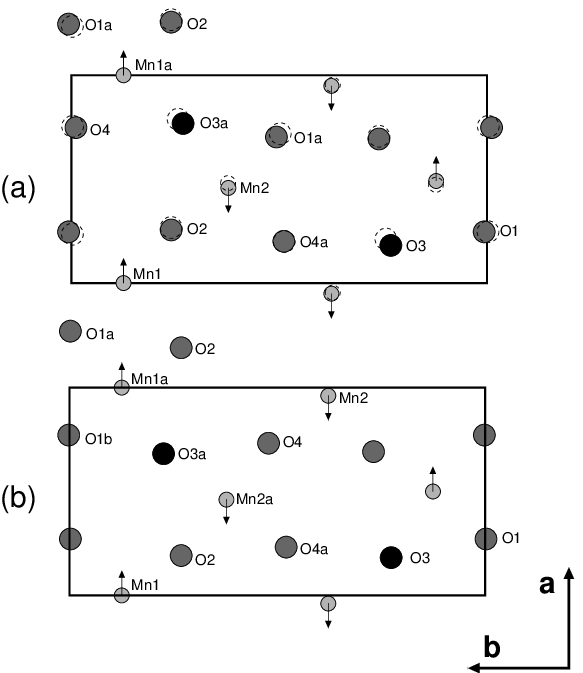}
\caption{$\bf{ab}$ plane atomic positions for \LaCa shown to scale. (a) \pntom structure. Mn ion positions in the ZP state (Table \ref{tab:tab1}) are indicated by small circles, \oion ions by shaded circles and \ooneion ions in the ZP state by a filled circle. Mn and O ion positions in the checkerboard state (Table \ref{tab:tab1}) are indicated by small and large dashed circles. (b) \ptonm ZP structure (Table \ref{tab:tab2}). Directions of transverse displacements of Mn ions from their ideal perovskite positions are indicated by arrows. Labels on ions refer to Tables \ref{tab:tab1} to \ref{tab:tab3}.} 
\label{fig:fig1}
\end{figure}

In this Letter we use hybrid Hartree-Fock/density functional theory \ab energy minimization calculations and show that the ground state CO crystal structure switches between CB and ZP as the percentage of Hartree-Fock (exact) exchange is varied. These two states are therefore similar in energy and the actual ground state may depend on specific A and B ions \cite{Rivadulla02}, or even sample preparation, history and ambient conditions. The predominant difference in charge populations of the two states is in charge on O ions rather than Mn ions and so CO may be associated with O ions rather than Mn ions.  Furthermore, we find that the lowest energy structure for either CO pattern has a different space group from those proposed for the CB \cite{Radaelli97} or ZP \cite{Daoud02} structures. The structure that we find has modulations of Mn and O ion positions about ideal perovskite positions which are parallel to $\bf{ab}$ planes and transverse to the $\bf{b}$ axis and are similar to those observed in layered \cite{Li01, Nagai02}and simple doped manganites \cite{Chen96, Radaelli97, Radaelli99, Kajimoto02, Loudon02, Pissas02}. A TA phonon softening mechanism for the CO transition would explain the observation of incommensurate order with wavevector $\bf{q}$ = (1-x)$\bf{b}$* \cite{Loudon04a} and may also be relevant to colossal magnetoresistance (CMR) phases of manganites where diffuse satellites in x-ray scattering have been observed \cite{Nelson01,Campbell01} and attributed to polarons with transverse displacements of ions \cite{Campbell01} similar to those in Fig. \ref{fig:fig1}(a). 

Structure optimizations were performed using the CRYSTAL program \cite{Crystal03}. A $\sqrt{2}$x2$\sqrt{2}$x2 unit cell (Fig. \ref{fig:fig1}) containing eight formula units with ferromagnetic (FM) order was used for optimizations while total energies of A and CE-type magnetically ordered structures were compared using $2\sqrt{2}$x2$\sqrt{2}$x2 unit cells. A single force evaluation was carried out for CE-type magnetic order for a structure which had been optimized with FM magnetic order in order to estimate the importance of magnetoelastic effects on the crystal structure; these were found to be small. No simultaneous relaxation of unit cell dimensions was performed as the calculations are expensive in computer time; unit cell dimensions were taken from experiment \cite{Radaelli97} (a = 5.4763 $\AA$, b = 10.8932 $\AA$, c = 7.5247 $\AA$). Structure optimizations were performed for 60\%, 80\% and 100\% exact exchange (the latter is simply an unrestricted Hartree-Fock  (UHF) calculation); both 60\% and 80\% exact exchange calculations resulted in CB CO for all space groups investigated while 100\% exact exchange resulted in ZP CO structures.  Initial atomic configurations were taken from Table II in Ref. [\onlinecite{Radaelli97}] (\ptom symmetry) or were generated by hand. As noted by Daoud-Aladine and coworkers \cite{Daoud02}, the isotropy subgroups of the parent high temperature \pbnm phase that have a doubled unit cell along the $\bf{b}$ axis are: \pemp, \ptonmp, \ptomp, \ptobp, \ptonb and \pbp. The relevant isotropy subgroups for ZP CO are \ptonm and \pem and for CB CO they are \ptom \cite{Radaelli97} and \pemp. 

Hessian matrices for energy minimized CB or ZP CO structures with these space groups had at least one negative eigenvalue, which indicates that they are saddle points on the potential energy surface. A further energy minimization was performed using \poto symmetry and 100\% exact exchange.  The screw axis parallel to the $\bf{b}$ axis naturally incorporates transverse Mn displacements found in experiment. All Hessian matrix eigenvalues were positive for this ZP CO structure and it was lower in energy than the other structures (\pem +7 meV/Mn ion; \ptonm +14 meV/Mn ion). A structure optimization with 60\% exact exchange and \pntom symmetry resulted in a stable CB CO structure which was 31 meV/Mn ion lower than the optimized structure with \ptom symmetry. Fractional coordinates for the lowest energy CB and ZP CO structures found, which have \pntom symmetry, are given in Table \ref{tab:tab1}. Fractional coordinates for the \LaCa ZP CO structure are compared to those determined by neutron scattering for \PrCasix \cite{Daoud02} in Table \ref{tab:tab2}. 

\begin{table}[ht!]
\caption{\label{tab:tab1}Fractional coordinates for \LaCa with \pntom symmetry from \ab total energy minimization. Wyckoff positions are given in the second column. Coordinates on the left were determined using 100\% exact exchange and those on the right using 60\% exact exchange. Both structures have the (x,y) coordinates of the Mn1 ion in common.}
\begin{ruledtabular}
\begin{tabular}{lccccccc}
   &   &  &  100\% & & & 60\%  \\
Atom &Wyck.  &x  &y  &z   &x  &y   &z\\
\hline
Ca1 &2a  &0.4773 &0.8866 &0.0000 &0.4861 &0.8866 &0.0000 \\
La2 &2a  &0.4652 &0.3732 &0.0000 &0.4514 &0.3673 &0.0000 \\
La3 &2a  &0.9760 &0.1226 &0.5000 &0.9743 &0.1261 &0.5000 \\
Ca4 &2a  &0.9789 &0.5985 &0.5000 &0.9779 &0.6057 &0.5000 \\
Mn1 &4b  &0.0000 &0.8746 &0.2445 &0.0000 &0.8713 &0.2478 \\
Mn2 &4b  &0.4578 &0.6224 &0.2534 &0.4811 &0.6231 &0.2535 \\
O1  &4b  &0.2463 &0.0069 &0.2109 &0.2352 &0.9977 &0.2109 \\
O2  &4b  &0.2556 &0.7600 &0.2112 &0.2676 &0.7600 &0.2144 \\
O3        &4b  &0.1813 &0.2315 &0.2042 &0.2118 &0.2433 &0.2146\\
O4        &4b  &0.7479 &0.9891 &0.2753 &0.7539 &0.9970 &0.2717\\
O1'       &2a  &0.4298 &0.1255 &0.5000 &0.4320 &0.1055 &0.5000\\
O2'       &2a  &0.5550 &0.1309 &0.0000 &0.5430 &0.1337 &0.0000\\
O3'       &2a  &0.0443 &0.3645 &0.5000 &0.0316 &0.3611 &0.5000\\
O4'       &2a  &0.9054 &0.3770 &0.0000 &0.9012 &0.3625 &0.0000\\
\end{tabular}        
\end{ruledtabular}
\end{table}

\begin{table}[ht!]
\caption{\label{tab:tab2}Fractional coordinates for \LaCa (LCMO) from \ab energy minimization with 100\% exact exchange and for \PrCasix (PCMO) from single crystal neutron diffraction \cite{Daoud02}, both with \ptonm symmetry. Wyckoff positions are given in the second column.}
\begin{ruledtabular}
\begin{tabular}{lcccccccr}
 &  &  &LCMO &  & &PCMO &\\
Atom &Wyck. &x &y &z &x &y &z\\
\hline
Ca1 &2a &0.4769 &0.9021 &0.0000 &0.5121 &0.8936 &0.0000\\
La2 &2a &0.4783 &0.3723 &0.5000 &0.4784 &0.3614 &0.5000\\
La3 &2a &0.9832 &0.1286 &0.0000 &0.9977 &0.1426 &0.0000\\
Ca4 &2a &0.9741 &0.6093 &0.5000 &0.9905 &0.6088 &0.5000\\
Mn1 &4b &0.0000 &0.8743 &0.2498 &0.0000 &0.8756 &0.2489\\
Mn2 &4b &0.9611 &0.3775 &0.7518 &0.9795 &0.3746 &0.7492\\
O1  &4b &0.2717 &0.9980 &0.2768 &0.3044 &0.9845 &0.2861\\
O2  &4b &0.6918 &0.2685 &0.7837 &0.7090 &0.2676 &0.7891\\
O3  &4b &0.1818 &0.2261 &0.7029 &0.2112 &0.2328 &0.7148\\
O4  &4b &0.7324 &0.5211 &0.2276 &0.7515 &0.5191 &0.2110\\
O1' &2a &0.4344 &0.1246 &0.0000 &0.4353 &0.1125 &0.0000\\
O2' &2a &0.5575 &0.1322 &0.5000 &0.5743 &0.1321 &0.5000\\
O3' &2a &0.0375 &0.3644 &0.0000 &0.0562 &0.3758 &0.0000\\
O4' &2a &0.9247 &0.3777 &0.5000 &0.9104 &0.3846 &0.5000\\
\end{tabular}
\end{ruledtabular}
\end{table}

Fig. \ref{fig:fig1}(a) shows atomic positions for nearly coplanar Mn and O ions projected onto the $\bf{ab}$ plane in the \pntom energy minimized structure. Positions of ions in the ZP CO structure are indicated by shaded or filled circles and positions of ions in the CB CO structure are indicated by dashed lines (Table \ref{tab:tab1}). The structures are coincident at the Mn1 ion positions and so this figure facilitates comparison of distortions in the $\bf{ab}$ plane. The main differences in ion position occur at the Mn2 and O3 positions to accommodate the switch between ZP and CO JT distortion patterns.
Transverse wavelike modulations of Mn and O ion positions about ideal perovskite positions can be identified in Fig. \ref{fig:fig1}(a). Modulations of either ion type have wavelength equivalent to the unit cell dimension along the $\bf{b}$ axis and are out of phase by $\pi/2$ and are similar to modulations of \Mnoct proposed in manganites with x $>$ 0.5 \cite{Nagai02} and x $<$ 0.5 \cite{Campbell01}. 

Positions of ions in the energy minimized \ptonm structure with ZP CO are shown in Fig. \ref{fig:fig1}(b). The main differences in atomic positions in the plane between this ZP CO structure and the \pntom ZP CO structure occur at the O2 and O3 positions. Fractional coordinates for the \ptonm structures for both \LaCa (this work) and for \PrCasix \cite{Daoud02} are compared in Table \ref{tab:tab2} and the agreement is remarkable. The starting guess for the \LaCa structure in \ab calculations was not the \PrCasix structure and the coordinates obtained via \ab calculations are not simply a relaxation of the \PrCasix structure. There are bonds of intermediate length along the ZP axis ranging from 1.99 to 2.09 $\AA$ in \ab calculations and from 1.98 to 2.05 $\AA$ in experiment \cite{Daoud02} (Table \ref{tab:tab3}). Bond valence sums \cite{Brown92} (calculated using R$_{o}$ and B values of 1.750 and 0.37 for all Mn-O bonds) for the ZP structures are close to 3.5 in both computed and experimental ZP structures, which is indicative of an intermediate valence. The Mn-\ooneionp-Mn bond angle is 160$^{o}$ or less in both \LaCa and \PrCasixp. 

\begin{figure}[h!]
\includegraphics[width=8.7cm,height=4.35cm]{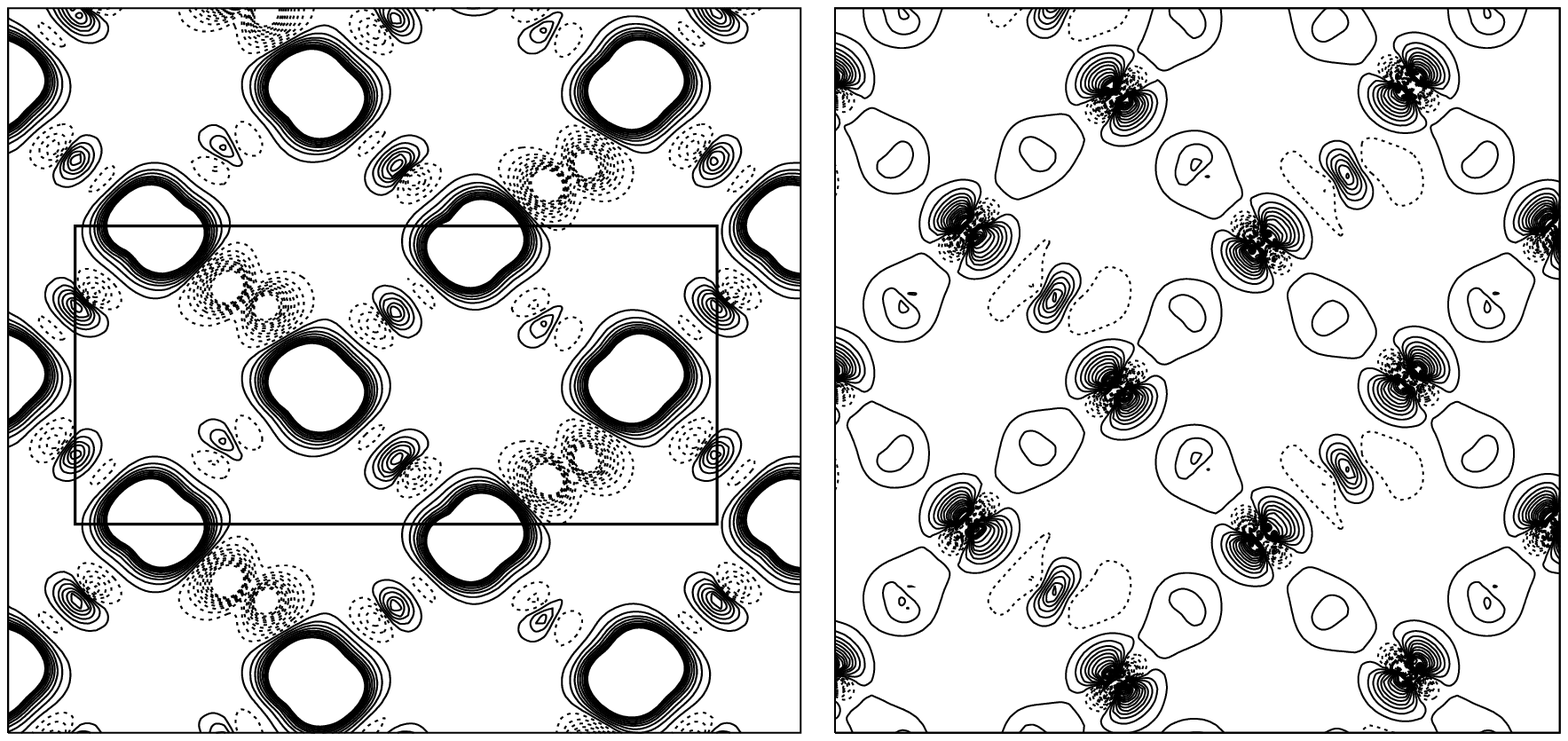}
\caption{Spin density and charge density difference for \LaCa in the ZP state with A-type magnetic order and \pntom symmetry obtained using 100\% exact exchange. (left) Spin density. The unit cell shown in Fig. \ref{fig:fig1}(a) is outlined. (right) Charge density difference.}
\label{fig:fig2}
\end{figure}

\begin{table}
\caption{\label{tab:tab3}Selected bond distances in \AA, bond angles for LCMO from \ab energy minimization and for PCMO from single crystal neutron diffraction \cite{Daoud02} and bond valence sums (BVS). Ion labels refer to atomic positions given in Fig. \ref{fig:fig1}. The bond angle given is the Mn1a-O3a-Mn2 angle at the center of the ZP or adjacent to a JT distorted Mn ion. The structure and space group are indicated at the top of each column.}
\begin{ruledtabular}
\begin{tabular}{lccccc}
Bond  & CB\footnotemark[1]   & ZP\footnotemark[2]   & Bond & ZP\footnotemark[3]   & ZP \footnotemark[4] \\
\hline
Mn1a-O1a        &1.92 &1.99 & Mn1a-O1a        &2.01        &2.05        \\
Mn1a-O4        &1.86 &1.87 & Mn1a-O1b        &1.86        &1.88        \\
Mn1a-O2        &1.91 &1.89 & Mn1a-O2        &1.91        &1.91        \\
Mn1a-O3a        &1.84 &2.05 & Mn1a-O3a        &2.08        &1.98        \\
Mn2-O2        &1.91 &1.89 & Mn2a-O2        &1.90        &1.90        \\
Mn2-O3a        &2.14 &2.09 & Mn2a-O3a        &2.09        &2.01        \\
Mn2-O1a        &1.92 &1.86 & Mn2a-O4        &1.88        &1.90        \\

Mn2-O4a        &2.15 &2.03 & Mn2a-O4a        &2.02        &2.03        \\
\hline
Bond angle  &160 &155  & &151 &159                    \\
BVS Mn1 & 3.28 & 3.60 & & 3.60 & 3.5 \footnotemark[4] \\
BVS Mn2 & 4.08 & 3.56 & & 3.53 & 3.5 \footnotemark[4] \\
\end{tabular}
\footnotemark[1]{\LaCa \pntom symmetry, 60\% exact exchange}\\
\footnotemark[2]{\LaCa \pntom symmetry, 100\% exact exchange}\\
\footnotemark[3]{\LaCa \ptonm symmetry, 100\% exact exchange }\\
\footnotemark[4]{\PrCasix \ptonm symmetry, expt. (Ref. \cite{Daoud02})}\\
\end{ruledtabular}
\end{table}

The spin density and charge density difference of the ZP CO structure are shown in Fig. \ref{fig:fig2}. Charge density difference plots are generated by subtracting densities of isolated \oion and \mmmm ions from the total charge density of the crystal structures and therefore show deformations of charge density at O ion sites and \eg orbital order at Mn ion sites. Each Mn ion in the ZP CO state has a \df configuration, \ooneion ions order in Mn(\dfp)-\ooneionp-Mn(\dfp) dimers (Zener polarons) in $\bf{ab}$ planes and the magnetic ground state is A-type \cite{Zheng03}. UHF calculations \cite{Zheng03} predict a magnetic moment of 0.7 \mub on the \ooneion ion in the center of each ZP, which is opposed to the moments of neighboring ZP Mn ions. On the other hand, cluster configuration interaction (CI) calculations \cite{Patterson04} show that the moment on \ooneion ions is much less than 0.7 \mubp. However they show that the charge on these ions is approximately -1.0e, in agreement with UHF calculations and that the ZP are strongly bound in a FM state. In both UHF and CI calculations the total magnetic moment on each ZP is 7/2 \mub in agreement with experiment \cite{Loudon02}. The reason for the discrepancy between spin distributions in CI and UHF calculations is simply that the UHF spin function for the Mn-\ooneion bond is $\alpha\beta$ whereas it should be $(\alpha\beta-\beta\alpha) /\sqrt{2}$. \dx and \dyy orbital order in the ZP state can clearly be seen in Fig. \ref{fig:fig2} in both spin density and charge density difference plots.

\begin{figure}[h!]
\includegraphics[width=8.7cm,height=4.35cm]{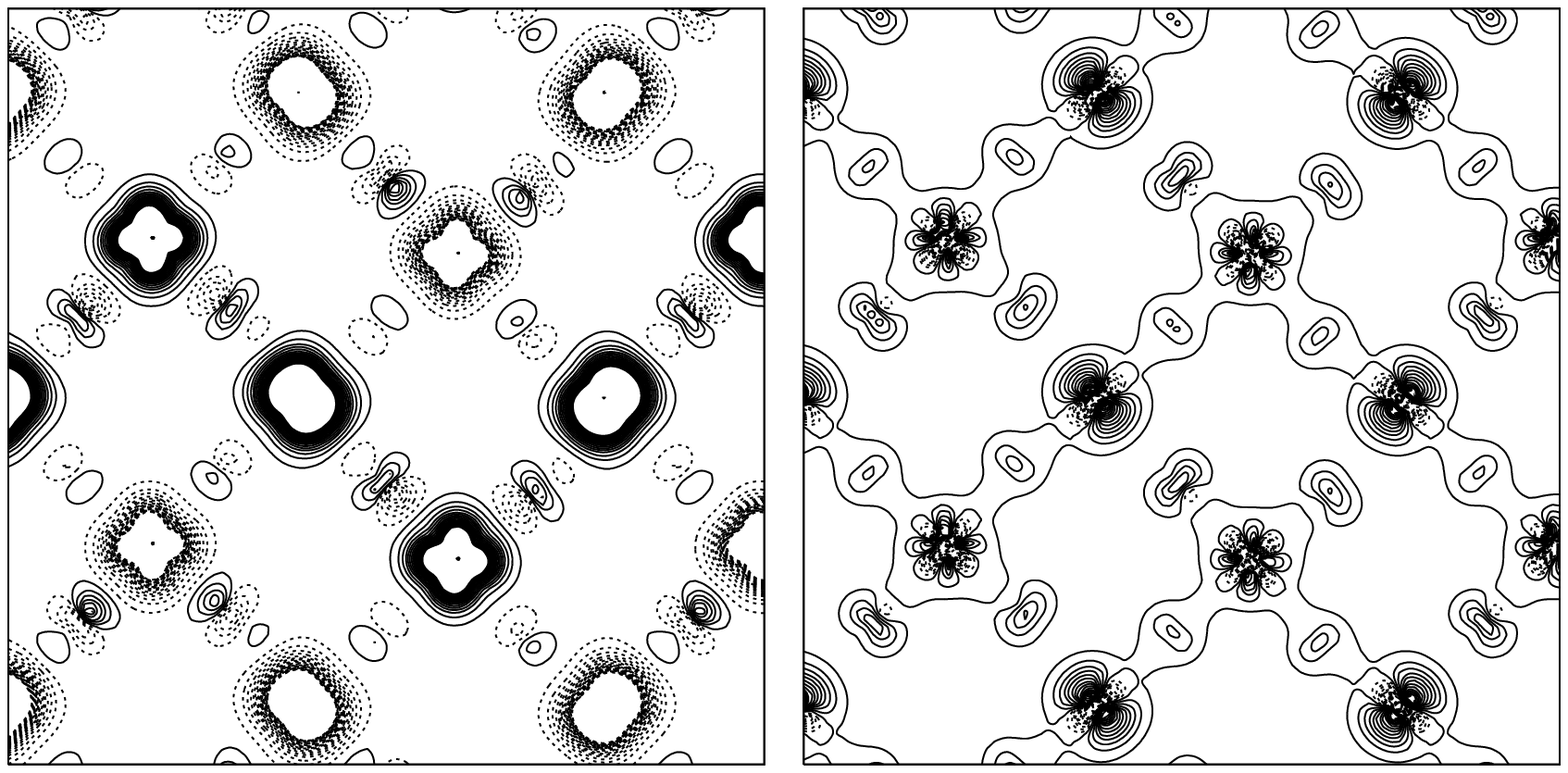}
\caption{Spin density and charge density difference for \LaCa in the checkerboard state with CE-type magnetic order and \pntom symmetry obtained using 60\% exact exchange. (left) Spin density. (right) Charge density difference.}
\label{fig:fig3}
\end{figure}

Bond lengths, the Mn1a-O3a-Mn2 bond angle and bond valence sums for the CB state with \pntom symmetry are given on the left in Table \ref{tab:tab3}. Spin densities and charge density differences are given in Fig. \ref{fig:fig3}. The ground state magnetic order for the CB CO state is CE-type, the magnetic order found experimentally in \LaCa \cite{Radaelli97}, and it consists of zig-zag FM chains as shown in Fig. \ref{fig:fig3}. Mn1 ions in the chains have short (1.85$\AA$) Mn-O bonds while JT distorted Mn2 ions have long Mn-O bonds (2.15$\AA$). These are, respectively, the lengths of Mn-O bonds in \Ca and \La \cite{Wollan55,Rodriguez98}. The magnetic moments on these ions are 3.23\mub (Mn1) and 3.85\mub (Mn2) and bond valence sums are 3.28 and 4.08. All of these are characteristic of conventional CO, although charges on Mn ions as measured by Mulliken populations are essentially identical, with values of +2.02 (Mn1) and +2.02 (Mn2). A small difference in Mn ion charge has been noted in several experimental papers \cite{Grenier04,Goff04} in apparently CO states. 

In summary, we have shown that the equilibrium structure for \LaCa which is predicted by hybrid Hartree-Fock/density functional theory depends on the percentage of exact exchange used in the calculation. The CB CO state consists of FM zig-zag chains in which corner (Mn1) ions have short bonds (1.85$\AA$) while JT distorted (Mn2) ions have long bonds (2.15$\AA$) to neighboring O ions. 
A preliminary calculation of the RXD spectrum using this structure is in reasonably good agreement with experimental RXD spectra for \PrCasix \cite{Grenier04priv}.  The ZP CO state consists of FM planes tiled with polarons which have Mn-O bonds of intermediate length along the polaron axis and this structure is in reasonable agreement with that found for \PrCasix by neutron scattering \cite{Daoud02}. Transformation from the CB to the ZP state requires rehybridization at Mn1 and O3 sites together with relatively minor displacements shown in Fig. \ref{fig:fig1}(a). Effective magnetic moments from magnetic susceptibility data for at least two mixed valence manganites display marked increases in effective magnetic moment on cooling below the CO transition temperature to 6.1 \cite{Prodi04} or 7.9 \mub \cite{Daoud02}. These magnitudes are similar to the ZP magnetic moment of 7 \mub. Dimerization of CE zig-zag chains to form a ZP state may produce this large effective moment. 

\begin{acknowledgments}
This work was supported by the Irish Higher Education Authority under the PRTLI-IITAC2 programme.  The author wishes to acknowledge discussions with P.G. Radaelli, A. Daoud-Aladine and S. Grenier.
\end{acknowledgments}

\bibliography{LaCaCBZP}

\begin{thebibliography}{28}
\expandafter\ifx\csname natexlab\endcsname\relax\def\natexlab#1{#1}\fi
\expandafter\ifx\csname bibnamefont\endcsname\relax
  \def\bibnamefont#1{#1}\fi
\expandafter\ifx\csname bibfnamefont\endcsname\relax
  \def\bibfnamefont#1{#1}\fi
\expandafter\ifx\csname citenamefont\endcsname\relax
  \def\citenamefont#1{#1}\fi
\expandafter\ifx\csname url\endcsname\relax
  \def\url#1{\texttt{#1}}\fi
\expandafter\ifx\csname urlprefix\endcsname\relax\def\urlprefix{URL }\fi
\providecommand{\bibinfo}[2]{#2}
\providecommand{\eprint}[2][]{\url{#2}}

\bibitem[{\citenamefont{Chen and Cheong}(1996)}]{Chen96}
\bibinfo{author}{\bibfnamefont{C.}~\bibnamefont{Chen}} \bibnamefont{and}
  \bibinfo{author}{\bibfnamefont{S.-W.} \bibnamefont{Cheong}},
  \bibinfo{journal}{Phys. Rev. Lett.} \textbf{\bibinfo{volume}{76}},
  \bibinfo{pages}{4042} (\bibinfo{year}{1996}).

\bibitem[{\citenamefont{Radaelli et~al.}(1997)\citenamefont{Radaelli, Cox,
  Marezio, and Cheong}}]{Radaelli97}
\bibinfo{author}{\bibfnamefont{P.~G.} \bibnamefont{Radaelli}},
  \bibinfo{author}{\bibfnamefont{D.~E.} \bibnamefont{Cox}},
  \bibinfo{author}{\bibfnamefont{M.}~\bibnamefont{Marezio}}, \bibnamefont{and}
  \bibinfo{author}{\bibfnamefont{S.-W.} \bibnamefont{Cheong}},
  \bibinfo{journal}{Phys. Rev. B} \textbf{\bibinfo{volume}{55}},
  \bibinfo{pages}{3015} (\bibinfo{year}{1997}).

\bibitem[{\citenamefont{Radaelli et~al.}(1999)\citenamefont{Radaelli, Cox,
  Capogna, Cheong, and Marezio}}]{Radaelli99}
\bibinfo{author}{\bibfnamefont{P.~G.} \bibnamefont{Radaelli}},
  \bibinfo{author}{\bibfnamefont{D.~E.} \bibnamefont{Cox}},
  \bibinfo{author}{\bibfnamefont{L.}~\bibnamefont{Capogna}},
  \bibinfo{author}{\bibfnamefont{S.-W.} \bibnamefont{Cheong}},
  \bibnamefont{and} \bibinfo{author}{\bibfnamefont{M.}~\bibnamefont{Marezio}},
  \bibinfo{journal}{Phys. Rev. B} \textbf{\bibinfo{volume}{59}},
  \bibinfo{pages}{14440} (\bibinfo{year}{1999}).

\bibitem[{\citenamefont{Li et~al.}(2001)\citenamefont{Li, Jin, and
  Zhao}}]{Li01}
\bibinfo{author}{\bibfnamefont{J.}~\bibnamefont{Li}},
  \bibinfo{author}{\bibfnamefont{C.}~\bibnamefont{Jin}}, \bibnamefont{and}
  \bibinfo{author}{\bibfnamefont{H.}~\bibnamefont{Zhao}},
  \bibinfo{journal}{Phys. Rev. B} \textbf{\bibinfo{volume}{64}},
  \bibinfo{pages}{R20405} (\bibinfo{year}{2001}).

\bibitem[{\citenamefont{Nagai et~al.}(2002)\citenamefont{Nagai, Kimura,
  Yamazaki, Asaka, Kimoto, Tokura, and Matsui}}]{Nagai02}
\bibinfo{author}{\bibfnamefont{T.}~\bibnamefont{Nagai}},
  \bibinfo{author}{\bibfnamefont{T.}~\bibnamefont{Kimura}},
  \bibinfo{author}{\bibfnamefont{A.}~\bibnamefont{Yamazaki}},
  \bibinfo{author}{\bibfnamefont{T.}~\bibnamefont{Asaka}},
  \bibinfo{author}{\bibfnamefont{K.}~\bibnamefont{Kimoto}},
  \bibinfo{author}{\bibfnamefont{Y.}~\bibnamefont{Tokura}}, \bibnamefont{and}
  \bibinfo{author}{\bibfnamefont{Y.}~\bibnamefont{Matsui}},
  \bibinfo{journal}{Phys. Rev. B} \textbf{\bibinfo{volume}{65}},
  \bibinfo{pages}{R60405} (\bibinfo{year}{2002}).

\bibitem[{\citenamefont{Kajimoto et~al.}(2002)\citenamefont{Kajimoto,
  Yoshizawa, Tomioka, and Tokura}}]{Kajimoto02}
\bibinfo{author}{\bibfnamefont{R.}~\bibnamefont{Kajimoto}},
  \bibinfo{author}{\bibfnamefont{H.}~\bibnamefont{Yoshizawa}},
  \bibinfo{author}{\bibfnamefont{Y.}~\bibnamefont{Tomioka}}, \bibnamefont{and}
  \bibinfo{author}{\bibfnamefont{Y.}~\bibnamefont{Tokura}},
  \bibinfo{journal}{Phys. Rev. B} \textbf{\bibinfo{volume}{66}},
  \bibinfo{pages}{R180402} (\bibinfo{year}{2002}).

\bibitem[{\citenamefont{Loudon et~al.}(2002)\citenamefont{Loudon, Mathur, and
  Midgley}}]{Loudon02}
\bibinfo{author}{\bibfnamefont{J.~C.} \bibnamefont{Loudon}},
  \bibinfo{author}{\bibfnamefont{N.~D.} \bibnamefont{Mathur}},
  \bibnamefont{and} \bibinfo{author}{\bibfnamefont{P.~A.}
  \bibnamefont{Midgley}}, \bibinfo{journal}{Nature}
  \textbf{\bibinfo{volume}{420}}, \bibinfo{pages}{797} (\bibinfo{year}{2002}).

\bibitem[{\citenamefont{Pissas and Kallias}()}]{Pissas02}
\bibinfo{author}{\bibfnamefont{M.}~\bibnamefont{Pissas}} \bibnamefont{and}
  \bibinfo{author}{\bibfnamefont{G.}~\bibnamefont{Kallias}},
  \eprint{cond-mat/0205410}.

\bibitem[{\citenamefont{Loudon et~al.}()\citenamefont{Loudon, Cox, Williams,
  Attfield, Littewood, Midgley, and Mathur}}]{Loudon04a}
\bibinfo{author}{\bibfnamefont{J.~C.} \bibnamefont{Loudon}},
  \bibinfo{author}{\bibfnamefont{S.}~\bibnamefont{Cox}},
  \bibinfo{author}{\bibfnamefont{A.~J.} \bibnamefont{Williams}},
  \bibinfo{author}{\bibfnamefont{J.~P.} \bibnamefont{Attfield}},
  \bibinfo{author}{\bibfnamefont{P.~B.} \bibnamefont{Littewood}},
  \bibinfo{author}{\bibfnamefont{P.~A.} \bibnamefont{Midgley}},
  \bibnamefont{and} \bibinfo{author}{\bibfnamefont{N.~D.}
  \bibnamefont{Mathur}}, \eprint{cond-mat/0308501}.

\bibitem[{\citenamefont{Wollan and Koehler}(1955)}]{Wollan55}
\bibinfo{author}{\bibfnamefont{E.~O.} \bibnamefont{Wollan}} \bibnamefont{and}
  \bibinfo{author}{\bibfnamefont{W.~C.} \bibnamefont{Koehler}},
  \bibinfo{journal}{Phys. Rev.} \textbf{\bibinfo{volume}{100}},
  \bibinfo{pages}{545} (\bibinfo{year}{1955}).

\bibitem[{\citenamefont{Goodenough}(1955)}]{Goodenough55}
\bibinfo{author}{\bibfnamefont{J.~B.} \bibnamefont{Goodenough}},
  \bibinfo{journal}{Phys. Rev.} \textbf{\bibinfo{volume}{100}},
  \bibinfo{pages}{564} (\bibinfo{year}{1955}).

\bibitem[{\citenamefont{Zheng and Patterson}(2003)}]{Zheng03}
\bibinfo{author}{\bibfnamefont{G.}~\bibnamefont{Zheng}} \bibnamefont{and}
  \bibinfo{author}{\bibfnamefont{C.~H.} \bibnamefont{Patterson}},
  \bibinfo{journal}{Phys. Rev. B} \textbf{\bibinfo{volume}{67}},
  \bibinfo{pages}{220404(R)} (\bibinfo{year}{2003}).

\bibitem[{\citenamefont{Ferrari et~al.}(2003)\citenamefont{Ferrari, Towler, and
  Littlewood}}]{Ferrari03}
\bibinfo{author}{\bibfnamefont{V.}~\bibnamefont{Ferrari}},
  \bibinfo{author}{\bibfnamefont{M.}~\bibnamefont{Towler}}, \bibnamefont{and}
  \bibinfo{author}{\bibfnamefont{P.}~\bibnamefont{Littlewood}},
  \bibinfo{journal}{Phys. Rev. Lett.} \textbf{\bibinfo{volume}{91}},
  \bibinfo{pages}{227202} (\bibinfo{year}{2003}).

\bibitem[{\citenamefont{Daoud-Aladine et~al.}(2002)\citenamefont{Daoud-Aladine,
  Rodriguez-Carvajal, Pinsard-Gaudart, Fernandez-Diaz, and
  Revcolevschi}}]{Daoud02}
\bibinfo{author}{\bibfnamefont{A.}~\bibnamefont{Daoud-Aladine}},
  \bibinfo{author}{\bibfnamefont{J.}~\bibnamefont{Rodriguez-Carvajal}},
  \bibinfo{author}{\bibfnamefont{L.}~\bibnamefont{Pinsard-Gaudart}},
  \bibinfo{author}{\bibfnamefont{M.~T.} \bibnamefont{Fernandez-Diaz}},
  \bibnamefont{and}
  \bibinfo{author}{\bibfnamefont{A.}~\bibnamefont{Revcolevschi}},
  \bibinfo{journal}{Phys. Rev. Lett.} \textbf{\bibinfo{volume}{89}},
  \bibinfo{pages}{97205} (\bibinfo{year}{2002}).

\bibitem[{\citenamefont{Grenier et~al.}(2004)\citenamefont{Grenier, Hill,
  Gibbs, Thomas, v.~Zimmermann, Nelson, Kiryukhin, Tokura, Tomioka, Casa
  et~al.}}]{Grenier04}
\bibinfo{author}{\bibfnamefont{S.}~\bibnamefont{Grenier}},
  \bibinfo{author}{\bibfnamefont{J.}~\bibnamefont{Hill}},
  \bibinfo{author}{\bibfnamefont{D.}~\bibnamefont{Gibbs}},
  \bibinfo{author}{\bibfnamefont{K.}~\bibnamefont{Thomas}},
  \bibinfo{author}{\bibfnamefont{M.}~\bibnamefont{v.~Zimmermann}},
  \bibinfo{author}{\bibfnamefont{C.}~\bibnamefont{Nelson}},
  \bibinfo{author}{\bibfnamefont{V.}~\bibnamefont{Kiryukhin}},
  \bibinfo{author}{\bibfnamefont{Y.}~\bibnamefont{Tokura}},
  \bibinfo{author}{\bibfnamefont{Y.}~\bibnamefont{Tomioka}},
  \bibinfo{author}{\bibfnamefont{D.}~\bibnamefont{Casa}}, \bibnamefont{et~al.},
  \bibinfo{journal}{Phys. Rev. B} \textbf{\bibinfo{volume}{69}},
  \bibinfo{pages}{134419} (\bibinfo{year}{2004}).

\bibitem[{\citenamefont{Goff and Attfield}(2004)}]{Goff04}
\bibinfo{author}{\bibfnamefont{R.~J.} \bibnamefont{Goff}} \bibnamefont{and}
  \bibinfo{author}{\bibfnamefont{J.~P.} \bibnamefont{Attfield}},
  \bibinfo{journal}{Phys. Rev. B} \textbf{\bibinfo{volume}{70}},
  \bibinfo{pages}{140404R} (\bibinfo{year}{2004}).

\bibitem[{\citenamefont{F.~Rivadulla and Goodenough}(2002)}]{Rivadulla02}
\bibinfo{author}{\bibfnamefont{J.-S.~Z.} \bibnamefont{F.~Rivadulla},
  \bibfnamefont{E.~Winkler}} \bibnamefont{and}
  \bibinfo{author}{\bibfnamefont{J.}~\bibnamefont{Goodenough}},
  \bibinfo{journal}{Phys. Rev. B} \textbf{\bibinfo{volume}{66}},
  \bibinfo{pages}{174432} (\bibinfo{year}{2002}).

\bibitem[{\citenamefont{Sub\'ias et~al.}(1997)\citenamefont{Sub\'ias, Garc\'ia,
  Proietti, and Blasco}}]{Subias97}
\bibinfo{author}{\bibfnamefont{G.}~\bibnamefont{Sub\'ias}},
  \bibinfo{author}{\bibfnamefont{J.}~\bibnamefont{Garc\'ia}},
  \bibinfo{author}{\bibfnamefont{M.}~\bibnamefont{Proietti}}, \bibnamefont{and}
  \bibinfo{author}{\bibfnamefont{J.}~\bibnamefont{Blasco}},
  \bibinfo{journal}{Phys. Rev. B} \textbf{\bibinfo{volume}{56}},
  \bibinfo{pages}{8183} (\bibinfo{year}{1997}).

\bibitem[{\citenamefont{Garc\'ia
  et~al.}(2001{\natexlab{a}})\citenamefont{Garc\'ia, S\'anchez, Sub\'ias, and
  Blasco}}]{Garcia01}
\bibinfo{author}{\bibfnamefont{J.}~\bibnamefont{Garc\'ia}},
  \bibinfo{author}{\bibfnamefont{M.~C.} \bibnamefont{S\'anchez}},
  \bibinfo{author}{\bibfnamefont{G.}~\bibnamefont{Sub\'ias}}, \bibnamefont{and}
  \bibinfo{author}{\bibfnamefont{J.}~\bibnamefont{Blasco}},
  \bibinfo{journal}{J. Phys. Condens. Matt.} \textbf{\bibinfo{volume}{13}},
  \bibinfo{pages}{3229} (\bibinfo{year}{2001}{\natexlab{a}}).

\bibitem[{\citenamefont{Garc\'ia
  et~al.}(2001{\natexlab{b}})\citenamefont{Garc\'ia, S\'anchez, Blasco,
  Sub\'ias, and Proietti}}]{Garcia01a}
\bibinfo{author}{\bibfnamefont{J.}~\bibnamefont{Garc\'ia}},
  \bibinfo{author}{\bibfnamefont{M.~C.} \bibnamefont{S\'anchez}},
  \bibinfo{author}{\bibfnamefont{J.}~\bibnamefont{Blasco}},
  \bibinfo{author}{\bibfnamefont{G.}~\bibnamefont{Sub\'ias}}, \bibnamefont{and}
  \bibinfo{author}{\bibfnamefont{M.~G.} \bibnamefont{Proietti}},
  \bibinfo{journal}{J. Phys. Condens. Matt.} \textbf{\bibinfo{volume}{13}},
  \bibinfo{pages}{3243} (\bibinfo{year}{2001}{\natexlab{b}}).

\bibitem[{\citenamefont{Nelson et~al.}(2001)\citenamefont{Nelson,
  v.~Zimmermann, Kim, Hill, Gibbs, Kiryukhin, Koo, Cheong, Casa, Keimer
  et~al.}}]{Nelson01}
\bibinfo{author}{\bibfnamefont{C.}~\bibnamefont{Nelson}},
  \bibinfo{author}{\bibfnamefont{M.}~\bibnamefont{v.~Zimmermann}},
  \bibinfo{author}{\bibfnamefont{Y.}~\bibnamefont{Kim}},
  \bibinfo{author}{\bibfnamefont{J.}~\bibnamefont{Hill}},
  \bibinfo{author}{\bibfnamefont{D.}~\bibnamefont{Gibbs}},
  \bibinfo{author}{\bibfnamefont{V.}~\bibnamefont{Kiryukhin}},
  \bibinfo{author}{\bibfnamefont{T.}~\bibnamefont{Koo}},
  \bibinfo{author}{\bibfnamefont{S.-W.} \bibnamefont{Cheong}},
  \bibinfo{author}{\bibfnamefont{D.}~\bibnamefont{Casa}},
  \bibinfo{author}{\bibfnamefont{B.}~\bibnamefont{Keimer}},
  \bibnamefont{et~al.}, \bibinfo{journal}{Phys. Rev. B}
  \textbf{\bibinfo{volume}{64}}, \bibinfo{pages}{174405}
  (\bibinfo{year}{2001}).

\bibitem[{\citenamefont{Campbell et~al.}(2001)\citenamefont{Campbell, Osborn,
  Argyriou, Vasiliu-Doloc, Mitchell, Sinha, Ruett, Ling, Islam, and
  Lynn}}]{Campbell01}
\bibinfo{author}{\bibfnamefont{B.}~\bibnamefont{Campbell}},
  \bibinfo{author}{\bibfnamefont{R.}~\bibnamefont{Osborn}},
  \bibinfo{author}{\bibfnamefont{D.}~\bibnamefont{Argyriou}},
  \bibinfo{author}{\bibfnamefont{L.}~\bibnamefont{Vasiliu-Doloc}},
  \bibinfo{author}{\bibfnamefont{J.}~\bibnamefont{Mitchell}},
  \bibinfo{author}{\bibfnamefont{S.}~\bibnamefont{Sinha}},
  \bibinfo{author}{\bibfnamefont{U.}~\bibnamefont{Ruett}},
  \bibinfo{author}{\bibfnamefont{C.}~\bibnamefont{Ling}},
  \bibinfo{author}{\bibfnamefont{Z.}~\bibnamefont{Islam}}, \bibnamefont{and}
  \bibinfo{author}{\bibfnamefont{J.}~\bibnamefont{Lynn}},
  \bibinfo{journal}{Phys. Rev. B} \textbf{\bibinfo{volume}{65}},
  \bibinfo{pages}{014427} (\bibinfo{year}{2001}).

\bibitem[{\citenamefont{Saunders et~al.}()\citenamefont{Saunders, Dovesi,
  Roetti, Caus\'a, Orlando, Zicovich-Wilson, Harrison, Doll, Civalleri, Bush
  et~al.}}]{Crystal03}
\bibinfo{author}{\bibfnamefont{V.~R.} \bibnamefont{Saunders}},
  \bibinfo{author}{\bibfnamefont{R.}~\bibnamefont{Dovesi}},
  \bibinfo{author}{\bibfnamefont{C.}~\bibnamefont{Roetti}},
  \bibinfo{author}{\bibfnamefont{M.}~\bibnamefont{Caus\'a}},
  \bibinfo{author}{\bibfnamefont{R.}~\bibnamefont{Orlando}},
  \bibinfo{author}{\bibfnamefont{C.~M.} \bibnamefont{Zicovich-Wilson}},
  \bibinfo{author}{\bibfnamefont{N.~M.} \bibnamefont{Harrison}},
  \bibinfo{author}{\bibfnamefont{K.}~\bibnamefont{Doll}},
  \bibinfo{author}{\bibfnamefont{B.}~\bibnamefont{Civalleri}},
  \bibinfo{author}{\bibfnamefont{I.}~\bibnamefont{Bush}}, \bibnamefont{et~al.},
  \eprint{Crystal03 User's Manual, University of Torino, Torino, 2003.
  (www.crystal.unito.it)}.

\bibitem[{\citenamefont{Brown}(1992)}]{Brown92}
\bibinfo{author}{\bibfnamefont{I.~D.} \bibnamefont{Brown}},
  \bibinfo{journal}{Acta Crystallogr. B} \textbf{\bibinfo{volume}{48}},
  \bibinfo{pages}{553} (\bibinfo{year}{1992}).

\bibitem[{\citenamefont{Patterson}()}]{Patterson04}
\bibinfo{author}{\bibfnamefont{C.~H.} \bibnamefont{Patterson}}, \eprint{Mol.
  Phys. (Accepted for publication)}.

\bibitem[{\citenamefont{Rodriguez-Carvajal
  et~al.}(1998)\citenamefont{Rodriguez-Carvajal, Hennion, Moussa, Moudden,
  Pinsard, and Revcolevschi}}]{Rodriguez98}
\bibinfo{author}{\bibfnamefont{J.}~\bibnamefont{Rodriguez-Carvajal}},
  \bibinfo{author}{\bibfnamefont{M.}~\bibnamefont{Hennion}},
  \bibinfo{author}{\bibfnamefont{F.}~\bibnamefont{Moussa}},
  \bibinfo{author}{\bibfnamefont{A.~H.} \bibnamefont{Moudden}},
  \bibinfo{author}{\bibfnamefont{L.}~\bibnamefont{Pinsard}}, \bibnamefont{and}
  \bibinfo{author}{\bibfnamefont{A.}~\bibnamefont{Revcolevschi}},
  \bibinfo{journal}{Phys. Rev. B} \textbf{\bibinfo{volume}{57}},
  \bibinfo{pages}{R3189} (\bibinfo{year}{1998}).

\bibitem[{\citenamefont{Grenier}()}]{Grenier04priv}
\bibinfo{author}{\bibfnamefont{S.}~\bibnamefont{Grenier}}, \eprint{Private
  communication.}

\bibitem[{\citenamefont{Prodi et~al.}(2004)\citenamefont{Prodi, Gilioli,
  Gauzzi, Licci, Marezio, Bolzoni, Huang, Santoro, and Lynn}}]{Prodi04}
\bibinfo{author}{\bibfnamefont{A.}~\bibnamefont{Prodi}},
  \bibinfo{author}{\bibfnamefont{E.}~\bibnamefont{Gilioli}},
  \bibinfo{author}{\bibfnamefont{A.}~\bibnamefont{Gauzzi}},
  \bibinfo{author}{\bibfnamefont{F.}~\bibnamefont{Licci}},
  \bibinfo{author}{\bibfnamefont{M.}~\bibnamefont{Marezio}},
  \bibinfo{author}{\bibfnamefont{F.}~\bibnamefont{Bolzoni}},
  \bibinfo{author}{\bibfnamefont{Q.}~\bibnamefont{Huang}},
  \bibinfo{author}{\bibfnamefont{A.}~\bibnamefont{Santoro}}, \bibnamefont{and}
  \bibinfo{author}{\bibfnamefont{J.~W.} \bibnamefont{Lynn}},
  \bibinfo{journal}{Nature Mater.} \textbf{\bibinfo{volume}{3}},
  \bibinfo{pages}{48} (\bibinfo{year}{2004}).

\end{thebibliography}

\end{document}